\pdfoutput=1
\documentclass[10pt,journal]{IEEEtran}
\IEEEoverridecommandlockouts
\usepackage{cite}
\usepackage{amsmath,amssymb,amsfonts}
\usepackage{algorithmic}
\usepackage{graphicx}
\usepackage{textcomp}
\usepackage{xcolor}

\usepackage{microtype}
\usepackage{subfigure}
\usepackage{booktabs}
\usepackage{hyperref}
\usepackage{array}
\usepackage{multirow}
\usepackage{algorithm}
\usepackage{algorithmic}
\usepackage{comment}
\usepackage{wrapfig}
\usepackage{soul}

\newcommand{\USWIM}{\emph{U-SWIM}}

\newcommand{\ydel}[1]{} 
\newcommand{\yadd}[1]{#1}
\newcommand{\yrpl}[2]{\ydel{#1}\yadd{#2}}

\def\BibTeX{{\rm B\kern-.05em{\sc i\kern-.025em b}\kern-.08em
    T\kern-.1667em\lower.7ex\hbox{E}\kern-.125emX}}
\begin{document}


\title{U-SWIM: Universal Selective Write-Verify for Computing-in-Memory Neural Accelerators
}



\author{Zheyu~Yan,~\IEEEmembership{Student Member,~IEEE,}
        Xiaobo~Sharon~Hu,~\IEEEmembership{Fellow,~IEEE,}
        and~Yiyu~Shi,~\IEEEmembership{Senior~Member,~IEEE}
\thanks{Z. Yan, X. S. Hu, and Y. Shi are with the Department
of Computer Science and Engineering, University of Notre Dame, Notre Dame, IN, 46556 USA 
(e-mail: \{zyan2, shu, yshi4\}@nd.edu).}
\thanks{Please address comments to zyan2@nd.edu and yshi4@nd.edu.}
}

\markboth{IEEE Transactions on Computer Aided Design of Integrated Circuits \& Systems,~Vol.~1, No.~1, April~2023}%
{Jia \MakeLowercase{\textit{et al.}}: A Sample Article Using IEEEtran.cls for IEEE Journals}

\maketitle

\begin{abstract}

Architectures that incorporate Computing-in-Memory (CiM) using emerging non-volatile memory (NVM) devices have become strong contenders for deep neural network (DNN) acceleration due to their impressive energy efficiency. Yet, a significant challenge arises when using these emerging devices: they can show substantial variations during the weight-mapping process. This can severely impact DNN accuracy if not mitigated. A widely accepted remedy for imperfect weight mapping is the iterative write-verify approach, which involves \yrpl{checking}{verifying} conductance \yadd{values} and \yrpl{making adjustments}{adjusting devices} \yrpl{as required}{if needed}. In all existing publications, this procedure is applied to every individual device, resulting in a significant programming time overhead. In our research, we illustrate that only a \yrpl{selective}{small} fraction of weights need this write-verify treatment for the corresponding devices \yrpl{to preserve DNN accuracy}{and the DNN accuracy can be preserved}, yielding a notable programming acceleration. Building on this, we introduce \USWIM, a novel method \yrpl{driven by}{based on} the second derivative. It leverages a single iteration of forward and backpropagation to pinpoint the weights demanding write-verify. Through extensive tests on diverse DNN designs and datasets, \USWIM~manifests up to a 10x programming acceleration against the traditional exhaustive write-verify method, all while maintaining a similar accuracy level. Furthermore, compared to our earlier {\em SWIM} technique, \USWIM~excels, showing a 7x speedup when dealing with devices exhibiting non-uniform variations.

\end{abstract}

\section{Introductions}

Deep Neural Networks (DNNs) have surpassed human capabilities in various perception tasks such as image recognition, object detection, and speech recognition~\cite{liang2022variational, liang2021can, wang2021exploration}. Although the potential for deploying DNNs on edge devices like cars, smartphones, and intelligent sensors is promising, the limited computational capabilities and power constraints of these devices pose challenges for executing resource-intensive DNNs through CPUs or GPUs~\cite{sheng2022larger, sheng2023toward, sheng2023muffin}.

Non-volatile Computing-in-Memory (nvCiM) DNN accelerators~\cite{shafiee2016isaac, jiang2020device} offer a compelling alternative for edge computing applications by reducing data movement via in-place weight data access approach~\cite{sze2017efficient}. Utilizing next-generation non-volatile memory (NVM) technologies like resistive random-access memories (RRAMs), Magnetoresistive random-access memories (MRAMs), ferroelectric field-effect transistors (FeFETs), and phase-change memories (PCMs), nvCiM provides enhanced energy efficiency and greater memory density compared to conventional MOSFET-based architectures~\cite{chen2016eyeriss, yang2020co}.
Nonetheless, NVM technologies encounter various imperfections, particularly variations between devices stemming from manufacturing flaws, as well as inconsistencies across cycles attributable to the devices' random characteristics. Without adequate countermeasures, the weights actually mapped to these devices could deviate markedly from their intended values, causing significant degradation in performance.

Various methods have been proposed to address these challenges. Techniques like noise-aware training~\cite{jiang2020device, yan2022computing} and uncertainty-aware neural architecture search~\cite{yan2020single, yan2021uncertainty, yan2022radars} aim to improve the DNN robustness against device variations, they are not cost-efficient due to the necessity of retraining the DNNs from scratch. This makes it hard to utilize existing pre-trained models. On the other hand, on-chip in-situ training~\cite{yao2020fully} fine-tunes the DNNs directly once they are mapped onto nvCiM platforms, alleviating weight variation concerns. Although this method is clearly effective, it not only necessitates additional hardware to enable backpropagation and weight adjustments but also demands iterative training, involving multiple write cycles for each weight update, which can be time-consuming and harmful to the hardware due to limited write cycles for NVM devices.

Alternatively, a common approach to address these issues is the write-verify method. This method entails a repetitive cycle of writing and reading (verifying) pulses to confirm that the programmed weights differ from the target values by only a permissible amount. With this technique, the weight's deviation from its ideal value can be minimized to less than 3\%, and any reduction in DNN accuracy is confined to under 0.5\%~\cite{shim2020two}. However, the write-verify method is a time-consuming and labor-intensive procedure, requiring individual write-verify cycles for each weight. For example, the time needed to program a ResNet-18 for CIFAR-10 on an nvCiM platform can exceed a week~\cite{shim2020two}. Considering that the programming time scales linearly with the number of DNN parameters, and given that many state-of-the-art models have far more weights than a ResNet-18, it raises the question: \textbf{Is it absolutely necessary to write-verify each individual weight when mapping a DNN to an nvCiM platform?}

In this work, we demonstrate that the unambiguous response to the question previously raised is a resounding \textbf{NO}. It turns out that only a small subset of weights actually needs to undergo write-verify cycles to attain a DNN accuracy nearly matching that of a perfect mapping. As a result, we can drastically cut down the time required for programming on nvCiM platforms. To this end, we introduce \underline{U}niversal \underline{S}elective \underline{W}rite-verify for computing-\underline{I}n-\underline{M}emory neural accelerators (\USWIM). This approach diverges from the conventional write-verify method, which mandates write-verify operations for each and every weight. Inspired by~\cite{lecun1989optimal}, \USWIM~utilizes a metric based on weight second derivatives as a guide to pinpoint and focus on a particular set of \emph{sensitive} weights for the write-verify process. Additionally, in light of the impracticality of directly calculating these second derivatives using the finite difference method, we introduce a one-pass forward and backward propagation scheme similar to gradient computation to extract this second derivative information. Our experimental results on MNIST, CIFAR-10, and Tiny ImageNet indicate that \USWIM~can accelerate the programming process by up to 10x, 5x, and 9x, respectively, compared to the traditional exhaustive write-verify technique, a magnitude-based selective write-verify heuristic, and the state-of-the-art in-situ training methods.
Additionally, the proposed \USWIM~method achieves up to 7x programming speedup in comparison to our previously proposed selective write-verify method {\em SWIM}~\cite{yan2022swim} when addressing devices with non-uniform variations. This improvement stems from the fact that the previous method did not adequately support these types of devices. To the best of our knowledge, this is the first study that introduces and validates the concept of selective write-verify for programming nvCiM neural accelerators across various device types.
\section{Related Works}

\subsection{Crossbar Array-based Computing Engine}\label{sec:2.1}
\begin{figure}[ht]
\begin{center}
\centerline{\includegraphics[trim=0 150 550 0, clip, width=0.5\linewidth] {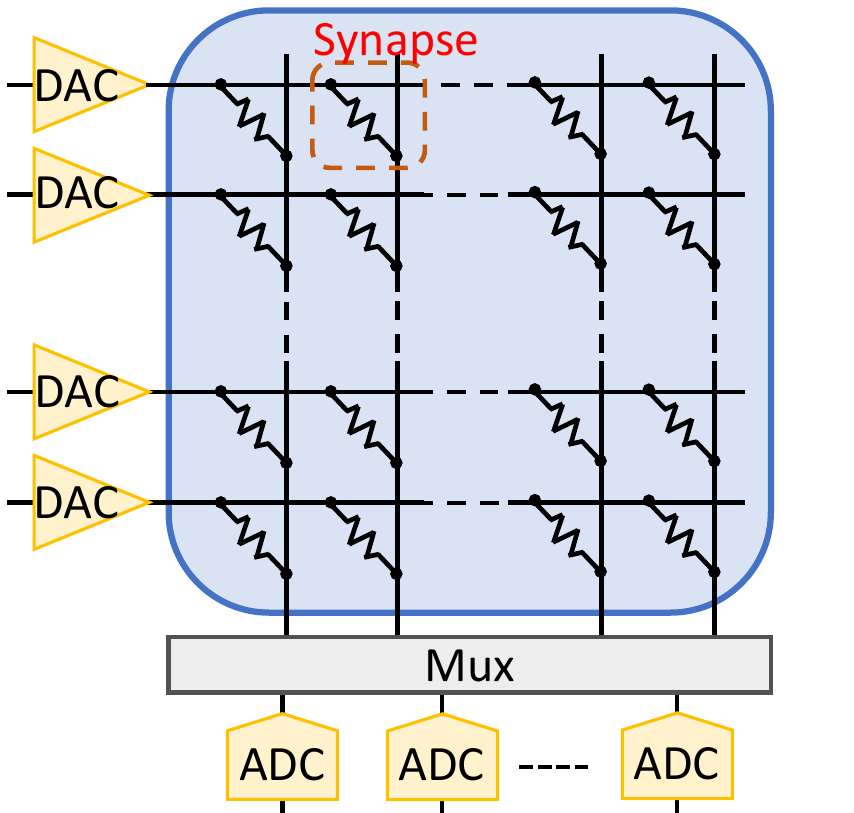}}
\caption{
Diagram of the crossbar array architecture: Inputs are introduced horizontally and multiplied by weights stored in the NVM devices at each intersection. The products are accumulated vertically to produce the final output.}
\label{fig:crossbar}
\end{center}
\end{figure}

The core building block of nvCiM DNN accelerators is the crossbar array architecture. This array serves as the computational unit for carrying out matrix-vector multiplication. In this array, matrix values, which are DNN weights if used in an nvCiM accelerator, are stored at the cross-points of vertical and horizontal lines through resistive emerging technologies like RRAMs, MRAMs, FeFETs, and PCMs. Simultaneously, vector values, which are DNN layer inputs if used in an nvCiM accelerator, flow along the horizontal lines of the array. Operations within the crossbar take place in the analog domain. However, for other essential DNN functions like pooling and non-linear activations, supplemental digital circuits are required. Consequently, digital-to-analog and analog-to-digital converters are employed to facilitate communication between these different components.

Crossbar arrays made of resistive devices are susceptible to a variety of \yrpl{fluctuations}{variations} and noise, with spatial and temporal variations being the main contributors. Spatial variations, often due to imperfections in the fabrication process, show both local and global correlations. Additionally, temporal variations occur owing to the inherent randomness in the material properties of NVM devices. These \yadd{issues will} result in \yrpl{inconsistent}{variations in} conductance levels when a device is programmed at different times. Such \yrpl{time-dependent}{temporal} variations are generally independent \yrpl{of each device}{across different devices} but can also be influenced by the targeted programmed value, as cited in prior research~\cite{feinberg2018making}. In this work, we \yrpl{center}{focus} our attention on assessing how these temporal variations during the programming phase impact the performance of DNNs. \yadd{Specifically,} these variations can lead to actual resistance levels in the devices deviating from what was initially expected, \yadd{Thus affecting the DNN weight values}. Our proposed methodology is flexible and can be adapted to account for other types of variations with some modifications.

\subsection{Handling Variations in Weight Mapping}
Various strategies have been proposed to mitigate the challenge of device variations in nvCiM DNN accelerators. Here, we briefly discuss two widely used strategies that obviate the need for re-training models from scratch.

One such strategy is on-chip in-situ training. This technique directly adjusts a pre-existing DNN on the nvCiM hardware to regain its performance, often within a limited number of iterations. This is particularly effective when the variations in the device are relatively minor. During this training process, both the forward and backward passes are conducted on the chip, with the device's variations influencing both passes. Voltage pulses are then applied to each corresponding device to update the weight, based on the weight's gradient. This method demands extra hardware resources and can be time-intensive, given the multiple write cycles needed for each weight. Moreover, it can be harmful to the NVM devices because these devices suffer from stuck-at-fault when written too many times. Some recent research suggests fine-tuning only the densely connected layers of the DNN model, although its efficacy on larger models is yet to be proved. \yadd{On the other hand, hardware-aware training~\cite{rasch2023hardware, yan2023improving} mimics this noisy inference process on CPU/GPU-based training platforms to improve the overall robustness of DNN models under the influence of device variations.}

The alternative strategy is the write-verify technique~\cite{shim2020two}. In this method, each NVM device is initially set using a predefined pulse sequence. The device's conductance is then read out (verified) to determine if it falls within an acceptable range of the targeted value (\emph{i.e.}, if its value is accurate). If not, another corrective pulse sequence is applied to bring the device's conductance closer to the desired level. This loop continues until the actual and targeted values are sufficiently close to each other. Typically, this process requires several iterations. \yadd{Various methods have been proposed to accelerate this process, including methods using gradient-based techniques~\cite{buchel2022gradient} and through discrete weight optimization~\cite{mackin2022optimised}.} Given that this process is done individually for each weight as it is mapped to a device, the time required for write-verify becomes considerably greater compared to methods that allow for parallel writes without verification.

\section{U-SWIM Framework}

\subsection{Overview of U-SWIM}

\begin{figure}[t]
\centering
\subfigure[]{
    \includegraphics[trim= 0 0 0 0, clip, width=0.45\linewidth]{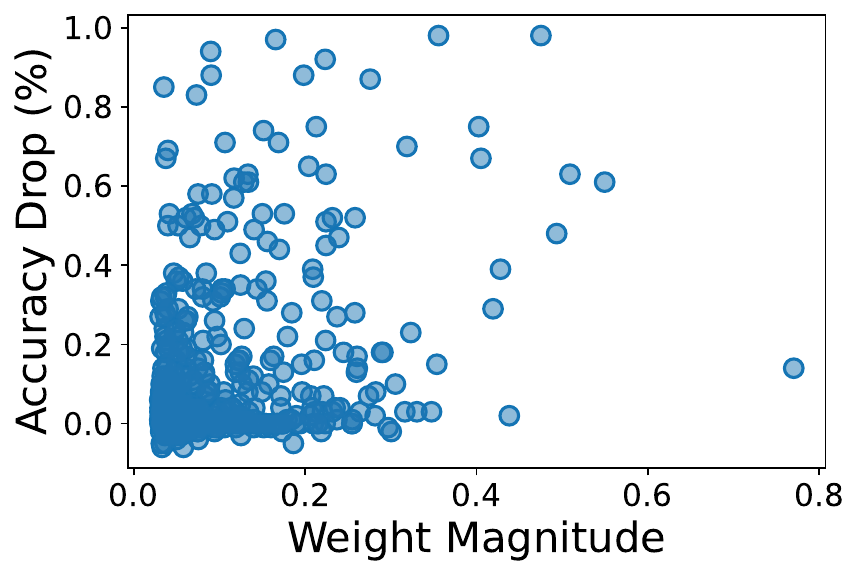}
    \label{fig:mag}
}
\subfigure[]{
    \includegraphics[trim= 0 0 0 0, clip, width=0.45\linewidth]{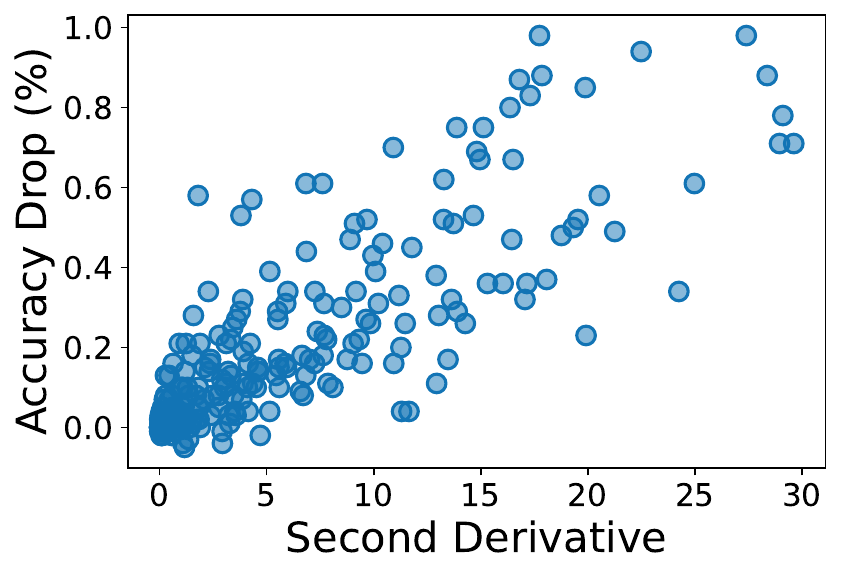}
    \label{fig:second}
}

\caption{
Effect of device variation-induced weight perturbation on LeNet DNN architecture targeting MNIST dataset: (a) Relationship between accuracy drop and weight magnitude, showing minimal correlation. (b) Correlation between accuracy drop and each weight's second derivative, indicating a strong correlation.
}
\label{fig:correlation}
\end{figure}

In this manuscript, instead of adhering to the conventional approach seen in existing works, which perform write-verify for every single weight of a DNN, we establish and investigate solutions for the following question.

\noindent {\bf Selective Write-Verify}: Given a DNN architecture with weights $\mathbf{W_0}$ and accuracy $A$. Also given a maximum acceptable accuracy drop $\delta A$. Identify the smallest subset $\mathbf{W_s}\subseteq \mathbf{W_0}$ so that, when mapping the DNN to nvCiM platforms, by only writing-verifying weights in $\mathbf{W_s}$, the deployed network can have an accuracy no less than $A - \delta A$. 

We propose a test-and-trial-based framework for this task. Specifically, A notable characteristic of NVM devices is that the reading process is considerably faster than writing, particularly for RRAMs and FeFETs~\cite{shafiee2016isaac}. This fast-read capability allows us to quickly evaluate the accuracy of a programmed DNN, making the read time negligible when compared to the time needed for write-verify. We exploit this characteristic to develop a heuristic method for tackling the selective write-verify challenge via iterative mapping, as outlined in Algorithm~\ref{alg:swim}. We begin by assessing the influence of device variation for each weight, in the initial weight matrix $\mathbf{W_0}$, on the DNN's accuracy. We term this impact as the weight's {\em sensitivity} for the remainder of this manuscript. Next, we rank all weights in descending order based on their {\em sensitivity} and apply write-verify in batches. Specifically, we group a fraction $p$ of the weights at a time (referred to as the programming granularity in Algorithm~\ref{alg:swim}) and continue this process until the drop in accuracy falls below $\delta A$. According to our preliminary experiments, a $p$ value that corresponds to $5\%$ of the total weight count offers a good balance, providing sufficient granularity for accuracy improvement while eliminating the need for overly frequent DNN accuracy evaluations. A pressing question that arises is how to assess the {\em sensitivity} of a weight. This topic will be addressed in the subsequent section.

\begin{algorithm}[h]
\caption{\USWIM~($\mathbf{W_0}$ , $\mathcal{Z}$,  $A$, $\delta A$, $\mathbf{D}$, $p$)}
\begin{algorithmic}[1]\label{alg:swim}
\STATE // INPUT: A trained DNN architecture $\mathcal{Z}$ with weights $\mathbf{W_0}$, original DNN accuracy $A$, the maximum accuracy drop allowed $\delta A$ after mapping to nvCiM, training dataset $\mathbf{D}$, and programming granularity $p$;

\STATE Program all weights in $\mathbf{W_0}$ based on their locations in $\mathcal{Z}$ to the nvCiM platform;
\STATE Calculate the \emph{sensitivity} of all the weights;
\STATE Sort $\mathbf{W_0}$ in the descending order of \emph{sensitivity}. 
\FOR{($i=0$; $i < (|\mathbf{W_0}|/p)$; $i++$)}
    \STATE Write-verify the weights $\mathbf{W_0}[i\times p+1: (i+1)\times p]$ based on their positions in $\mathcal{Z}$;
    \STATE Evaluate the accuracy $\tilde A$ of the mapped network on $\mathbf{D}$;
    \IF{$A-\tilde A \leq \delta A$}
        \STATE Break;
    \ENDIF
\ENDFOR
\end{algorithmic}
\end{algorithm}

\subsection{Sensitivity Analysis}\label{sec:sensitivity}

With the write-verify framework, our next objective is to identify a method for evaluating the {\em sensitivity} of individual weights. One might intuitively assume that a weight's magnitude serves as a good indicator of its {\em sensitivity} because heuristics show that larger weights would have a greater impact on accuracy if perturbed. However, our preliminary studies contradict this assumption.
We applied the same additive Gaussian noise to each weight in the LeNet model, as suggested by \cite{yao2020fully}. We then assessed the resulting change in DNN accuracy for each perturbed weight, averaging the results over 100 Monte Carlo runs. Figure \ref{fig:mag} reveals that there's a very weak, if any, correlation between the magnitude of weights and the decrease in accuracy due to weight perturbations. This observation is further validated in the experimental section, where we demonstrate that a magnitude-based selection approach falls short of expectations. Below, we present a rigorous mathematical analysis, to establish an effective metric that can accurately reflect the {\em sensitivity} of a weight. 

In current DNN optimization frameworks, maximizing accuracy is closely linked to minimizing a loss function. Therefore, the influence of a weight's variation on the model's accuracy is strongly correlated with its impact on this loss function. Consequently, we opt to assess the {\em sensitivity} of individual weights based on their effect on the loss function.

For a DNN equipped with a specified labeled training dataset, the loss $f$ acts as a function of a vector $\mathbf w$ that comprises all the weights. Presuming the training has converged and the optimal weights pinpointed are $\mathbf{\tilde{w}}$. When there are minor weight variations around $\mathbf{\tilde{w}}$, that is, $\mathbf{w} = \mathbf{\tilde{w}}+\mathbf{\Delta w}$, a Taylor expansion can be applied to $f$ in this manner:
\begin{equation}
    f(\mathbf{w}) = f(\mathbf{\tilde{w}}) + \frac{\partial f}{\partial \mathbf{\tilde{w}}} \Delta\mathbf{w} + \frac{1}{2} \Delta\mathbf{w}^T \mathcal{H}(\mathbf{\tilde{w}}) \Delta\mathbf{w} + o(\Delta \mathbf{w}^{3}) \label{eq:1}
\end{equation}
where the concise notation $\frac{\partial f}{\partial \mathbf{\tilde{w}}}$ denotes $\frac{\partial f}{\partial \mathbf{w}}|_{\mathbf{w}=\mathbf{\tilde{w}}}$. A similar abbreviation will be consistently used in this manuscript. $\mathcal{H}(\mathbf{w})$ is the Hessian of $\mathbf{w}$ and is defined as:
\begin{equation}
    \mathcal{H}(\mathbf{w}) = \left[\begin{array}{ccc}
\dfrac{\partial^2 f}{\partial w_1^2} & \cdots & \dfrac{\partial^2 f}{\partial w_1 \partial w_n} \\
\vdots & \ddots & \vdots \\
\dfrac{\partial^2 f}{\partial w_n \partial w_1} & \cdots & \dfrac{\partial^2 f}{\partial w_{n}^2}
\end{array}\right]
\end{equation}
where $n$ is the total weight count, \emph{i.e.}, length of $\mathbf{w}$.

The average performance change of this DNN model under the influence of device variations can be represented by the expectation of current $f$ value subtracted by the $f$ value when the model is not affected by device variations, \emph{i.e.}, $E[\Delta f(\mathbf{w})] = E[f(\mathbf{w}) - f(\mathbf{\tilde{w}})]$. Referring to Eq.~\ref{eq:1}, the expanded expectation is
\begin{equation}
    \begin{aligned}
        E[\Delta f(\mathbf{w})] = & E[f(\mathbf{w}) - f(\mathbf{\tilde{w}})] \\ = & \frac{\partial f}{\partial \mathbf{\tilde{w}}} E[\Delta\mathbf{w}] \\
        & + \frac{1}{2} E[\Delta\mathbf{w}^T \mathcal{H}(\mathbf{\tilde{w}}) \Delta\mathbf{w}] \\ 
        & + E[o(\Delta \mathbf{w}^{3})] \label{eq:expect}
    \end{aligned}
\end{equation}

We can not directly utilize this representation to find sensitive weights because it relies on multiple correlated factors. Therefore, we need to simplify this representation by including several assumptions.

\noindent\textbf{Assumption 1}: The device value deviation caused by device variations is orders of magnitude smaller than the maximum possible device conductance.

This assumption holds for most device programming schemes except for the corner case where designers choose to use too many levels in one device, leading to an unrealistic signal-to-noise ratio (SNR). Note that we are referring to the maximum possible conductance value, not the actual device conductance, so programming the device to a lower conductance value does not violate this assumption.

\noindent\textbf{Assumption 2}: The device value deviation caused by device variations follows a distribution with zero mean.

This assumption holds for most devices. We will discuss the level of approximations in the corner cases where this assumption does not hold.

According to assumption 2, $E[\Delta w_i] = 0$. Based on Eq.~\ref{eq:1}, its first-order term (\emph{i.e.}, $\frac{\partial f}{\partial \mathbf{\tilde{w}}} E[\Delta\mathbf{w}]$) is then zero. If assumption 2 does not hold, this term is also very small and can thus be ignored because when the DNN models are trained to converge, the first-order derivatives of each weight are very close to zero (\emph{i.e.}, $\frac{\partial f}{\partial \mathbf{\tilde{w}}} \approx 0$). Thus, even if assumption 2 does not hold, this term remains very close to zero and can be ignored.
Thus, the change in the loss function $E[\Delta f(\mathbf{w})]$ brought about by the weight variation $\Delta \mathbf{w}$ around $\mathbf{\tilde{w}}$
can be expressed as 
\begin{equation}
    E[\Delta f(\mathbf{w})] \approx  \frac{1}{2} E[\Delta\mathbf{w}^T \mathcal{H}(\mathbf{\tilde{w}}) \Delta\mathbf{w}]  \label{eq:2}
\end{equation}
where we have ignored the higher-order terms according to assumption 1.  

To explore further simplification, we notice that Eq.~\ref{eq:2} can be expressed as 
\begin{align}
\begin{split}
    &E[\Delta f(\mathbf{w})]  \approx \frac{1}{2} E\left[\sum^{n}_{i=1}\sum^{n}_{j=1} \mathcal{H}_{ij} \Delta w_i \Delta w_j\right]\\
    &=  \frac{1}{2} E\left[\sum^{n}_{i=1} \mathcal{H}_{ii} (\Delta w_i)^2\right] + \frac{1}{2} E\left[\sum^{n}_{i\neq j} \mathcal{H}_{ij} \Delta w_i \Delta w_j\right]\\
    & = \frac{1}{2} \sum^{n}_{i=1} \mathcal{H}_{ii} E[(\Delta w_i)^2] + \frac{1}{2} \sum^{n}_{i\neq j} \mathcal{H}_{ij} E[\Delta w_i \Delta w_j] \label{eq:taylor}
\end{split}
\end{align}
where $\Delta w_i$ is the $i^{th}$ element of $\Delta \mathbf{w}$ and $\mathcal{H}_{ij}$ is the element in the $i^{th}$ row and $j^{th}$ column of $\mathcal{H(\mathbf{\tilde{w}})}$. To further simplify this representation, we include a new assumption that,

\noindent\textbf{Assumption 3}: The correlated factors contributing to the device value deviation, caused by device variations, are orders of magnitude smaller than the independent factors, and therefore the correlated factors can be ignored.

This assumption holds for most devices. We will discuss the level of approximations in the corner cases where this assumption does not hold.

According to assumption 3, we have
\begin{equation}
    E[\Delta w_i \Delta w_j] = \begin{cases}
            E[(\Delta w_i)^2], & if\  i = j\\
            E[(\Delta w_i)]\times E[(\Delta w_j)], &if\   i \neq j\\
            \end{cases}
\end{equation}
and thus Eq.~\ref{eq:taylor} can be further simplified as   
\begin{equation}
    \begin{aligned}
        E[\Delta f(\mathbf{w})]  & \approx \frac{1}{2} \sum^{n}_{i=1} \mathcal{H}_{ii} E[(\Delta w_i)^2] & \ \\
        & \ \ \ +\frac{1}{2} \sum^{n}_{i\neq j} \mathcal{H}_{ij} E[\Delta w_i]\times E[\Delta w_j] & \  \\
        & = \frac{1}{2} \sum^{n}_{i=1} \mathcal{H}_{ii} E[(\Delta w_i)^2] \\
        & = \frac{1}{2} \sum^{n}_{i=1}\frac{\partial^2 f}{\partial \tilde{w}_i^2} E[(\Delta w_i)^2]
    \label{eq:final}
    \end{aligned}
\end{equation}
where $\tilde{w}_i$ is the $i^{th}$ element of $\mathbf{\tilde{w}}$. In the corner cases where the device correlation is stronger than the independent factor, this approximation is still acceptable. This is because the cross terms in the Hessian function in a DNN model are also orders of magnitudes smaller than the diagonal terms, which has been verified by previous research~\cite{lecun1989optimal}.

Eq.~\ref{eq:final} indicates that to evaluate the influence of weight variation on loss, we only need to derive the second derivative of each weight, denoted as $\frac{\partial^2 f}{\partial \tilde{w}_i^2}$, and additionally consider the variation of weight deviation distribution represented by $E[(\Delta w_i)^2]$. When we perform write-verify on weight $\tilde{w}_i$, we are effectively minimizing its variation, $\Delta w_{i}$. Hence, it stands to reason that we should prioritize minimizing the variation for weights characterized by elevated values of the metric $\frac{\partial^2 f}{\partial \tilde{w}_i^2} E[(\Delta w_i)^2]$. To put it simply, this metric serves as an effective {\em sensitivity} metric for \USWIM. The potency of this metric is reaffirmed in Fig.~\ref{fig:second}. Under conditions identical to those in Fig.~\ref{fig:mag}, a strong correlation emerges between the drop in accuracy after perturbing a weight and the introduced sensitivity metric of that particular weight, with a Pearson Correlation Coefficient \yadd{(PCC)} of 0.83, \yadd{which is significantly higher than using weight magnitude as a sensitivity metric, whose PCC is only 0.45. By observing the two figures, we can also conclude that both \USWIM~and the magnitude-based metric is able to identify some sensitive weights, but the magnitude-based method falls short in detecting those extremely sensitive ones (\emph{i.e.}, those with very high value in the Y-axis is not assigned high values in the X-axis.)}.

Finally, when two weights have the same $\frac{\partial^2 f}{\partial \tilde{w}_i^2} E[(\Delta w_i)^2]$, we use their magnitudes as the tie-breaker: the larger one will have a higher priority.

\subsection{Second Derivative Calculation}
When using $\frac{\partial^2 f}{\partial \tilde{w}_i^2} E[(\Delta w_i)^2]$ as a sensitivity metric, $E[(\Delta w_i)^2]$ can be easily derived from a device model (see Sect.~\ref{sect:model}). However, calculating the second derivative $\frac{\partial^2 f}{\partial \tilde{w}_i^2}$ of each weight \emph{w.r.t.} the model loss is not a simple task.

A direct approach to estimate the second derivative involves the finite difference method, which can be represented as:
\begin{equation}
   \frac{\partial^2 f}{\partial \tilde{w}_i^2}  \approx \frac{f(\tilde{w}_i+\Delta w)-2f(\tilde{w}_i)+f(\tilde{w}_i-\Delta w)}{(\Delta w)^2}
\end{equation}
where $\Delta w$ is a small positive value. However, to derive $f(\tilde{w}_i+\Delta w)$ and $f(\tilde{w}_i-\Delta w)$, it is necessary to run two forward propagation passes. This is done after substituting $\tilde{w}_i$ with $\tilde{w}_i+\Delta w$ and $\tilde{w}_i-\Delta w$ respectively. For a network consisting of a million weights, this translates to a staggering two million forward propagation runs.

Drawing inspiration from the efficient computation of all weight gradients through a single pass of forward and backpropagation, based on the chain rule, and the chain rule approximation for second derivatives outlined in~\cite{lecun1989optimal}, we introduce a method to similarly compute the second derivatives for all weights.

We'll begin by considering the last fully connected (FC) layer in a DNN model. The computations at this layer can be represented by the following equation:
\begin{equation}
 \begin{split}
     \mathbf{P} = g_a(\mathbf{I}),\  \mathbf{O} = \mathbf{W}\cdot \mathbf{P}
 \end{split}
 \end{equation}
where $g_a$ denotes the activation function of the previous layer, $\mathbf{I}$ is the input vector to that activation function, $\mathbf{W}$ is the weight matrix connecting the two layers, $\mathbf{P}$ is the output from the previous layer, and $\mathbf{O}$  is the output of the last layer. Note that we have excluded the activation function for the final layer, as it can be integrated into the loss function for simplicity.

Consider a loss function $f(\mathbf{O})$. Our aim is to compute the second derivatives with respect to the weights, $\frac{\partial^2 f}{\partial W_{ji}^2}$, and the inputs, $\frac{\partial^2 f}{\partial I_i^2}$. The former will serve as our {\em sensitivity} metric, while the latter will be utilized for further backpropagation to earlier layers. Given that $\mathbf O$ is a function of both $\mathbf{W}$ and $\mathbf{P}$, we can employ the chain rule for second derivatives as follows:
\begin{align}\label{eq:b_w}
\begin{split}
    \frac{\partial^2 f}{\partial W_{ji}^2} = \frac{\partial^2 f}{\partial O_{j}^2}\left(\frac{\partial O_{j}}{\partial W_{ji}}\right)^2+ \frac{\partial f}{\partial O_{j}}\frac{\partial^2 O_{j}}{\partial W_{ji}^2} = \frac{\partial^2 f}{\partial O_j^2}\times P_i^2 
\end{split}
\end{align}
where the second equality comes from the fact that $O_j$ is a linear function of $W_{ji}$ so $\frac{\partial^2 O_{j}}{\partial W_{ji}^2}=0$.
Similarly, we can get the second derivative of the input 
\begin{equation}
    \frac{\partial^2 f}{\partial I_i^2} = g_a'(P_i)^2 \sum_{j=1}^{|\mathbf{O}|} W_{ji}^2\frac{\partial^2 f}{\partial O_j^2} - g_a''(P_i) \frac{\partial f}{\partial I_i}
\end{equation}

Assuming ReLU is used as the activation function, the first derivative would be $g_a'(P_i) = sign(P_i) = sign(I_i)$ and the second derivative $g_a'' = 0$. In this case, the second derivatives of the input can be formulated as follows:
\begin{equation}\label{eq:b_a}
    \frac{\partial^2 f}{\partial I_i^2} = sign(I_i) \sum_{j=1}^{\mathbf{|O|}}W_{ji}^2 \frac{\partial^2 f}{\partial O_j^2}
\end{equation}

For max pooling layers, the backpropagation process cancels the derivatives of the deactivated inputs. That is, the second derivatives of these deactivated inputs become zero. In the case of architectures like ResNet that incorporate skip connections, the second derivatives from different branches are summed together, similar to the backpropagation process used for computing gradients. Convolution layers, average pooling, and batch normalization layers can be represented in the same form as fully connected (FC) layers, allowing their backpropagation to follow the same scheme used for FC layers.

Finally, to get these second derivatives of each weight, we also need to compute the second derivative of the loss functions with respect to the output of the DNN, \emph{i.e.}, $\frac{\partial^2 f}{\partial O_j^2}$.
For L2 loss, $\frac{\partial^2 f}{\partial O_j^2}=2$. For cross-entropy loss with softmax, 
\begin{equation}
\frac{\partial^2 f}{\partial O_j^2}     = \left(1 - \frac{O_j}{\sum_j \exp(O_j)}\right) \left(\frac{O_j}{\sum_j \exp(O_j)}\right)
\end{equation} 
We can then follow Eq.~\ref{eq:b_w} and Eq.~\ref{eq:b_a} to backpropagate layer by layer.

\noindent\yadd{\textbf{Overhead of second derivative calculation.}}

\yadd{The proposed second derivative calculation method requires similar computation and memory cost compared to gradient calculation.} \yrpl{Note that the}{Specifically, the} first-order gradient can be computed as 
\begin{align}\label{eq:g_w}
    \frac{\partial f}{\partial W_{ji}} &= \frac{\partial f}{\partial O_j} \times P_i \\
    \frac{\partial f}{\partial I_i} &= sign(I_i) \sum_{j=1}^{\mathbf{|O|}}W_{ji} \frac{\partial f}{\partial O_j}\label{eq:g_a}
\end{align}
Comparing Eq.~\ref{eq:g_w} and Eq.~\ref{eq:g_a} with Eq.~\ref{eq:b_w} and Eq.~\ref{eq:b_a}, it \yrpl{becomes}{is} evident that calculating the second derivative only involves an additional multiplication operation. The time required for this is negligible when compared to the time-intensive \yrpl{convolution operations carried out during forward propagation.}{matrix-vector multiplication in the same equation. For example, the computation overhead for calculating the second derivative for the LeNet model targeting the MNIST dataset is 0.2\%.} \yadd{Theoretically, there is also no memory overhead because gradient and second derivative calculation require saving the same data for backpropagation, which is the input of each layer.} With efficient implementation, the \USWIM~method's second derivative calculation requires approximately the same amount of time and memory as traditional gradient computations. Moreover, unlike the repeated gradient calculations needed for each iteration in gradient descent, the second derivative computation in \USWIM~is performed only once.

\section{Experimental Evalutaion}\label{sect:exp}
In this section, we first define the device variation model we use. Then, we perform a detailed study on the effectiveness of \USWIM~when using devices with uniform variations whose device value deviation distribution does not change with respect to device conductance values.
Firstly, we present an extensive analysis using the MNIST dataset to illustrate the superiority of \USWIM~over contemporary methods in various device variations. Subsequently, we employ the CIFAR-10 and Tiny ImageNet datasets to demonstrate their efficacy in more complex models.
We then conduct studies to demonstrate the effectiveness of \USWIM~when using devices with non-uniform variations whose device value deviation distribution changes with respect to device conductance values. Similar to previous studies, we first discuss the device models and then show the effectiveness of \USWIM~in the larger models. We did not perform experiments on MNIST because the difference between devices with uniform and non-uniform variations is small. Additionally, we conduct an ablation study to emphasize the importance of considering non-uniform device variation.

\subsection{Mapping and Impact of Device Variations}\label{sect:model}
This manuscript serves as a proof of concept, demonstrating the effectiveness of \USWIM~in addressing temporal variations during the programming phase. In this context, each device's variation is independent across devices but correlated with its own conductance. We employ a straightforward yet realistic model to capture this behavior.

For a quantized weight with a quantization precision of $M$ bits, let its desired value $\mathcal{W}_{des}$ be:
\begin{equation}
    \mathcal{W}_{des} = \sum_{i=0}^{M-1}{m_i \times 2^i}
\end{equation}
where $m_i$ represents the value of the desired weight at the $i^{th}$ bit. 
Additionally, we assume that the value programmed onto each device follows a Gaussian distribution of $\mathcal{N}(g_i, \sigma_i^2)$, with $g_i$ being the intended conductance value and $\sigma_i$ indicating the level of variability due to device variations.

When mapping the weight to nvCiM platforms, one weight value is mapped to one or multiple NVM devices. When using NVM device with $K$-bits, an $M$-bit weight can be mapped to $M/K$ devices\footnote{Wihtout loss of generality, we assume that M is a multiple of K.}, and the mapped value of the $i^{th}$ ($0\leq i\leq M/K-1$) device $g_i$ as:
\begin{equation}
    g_i = \mathcal{N}\left(\sum_{j=0}^{K -1}{ m_{i\times K + j} \times 2^{j}, \sigma_i^2 }\right)
\end{equation}
Note that weights with positive and negative values are mapped to two different arrays and they are mapped to the crossbar arrays according to their absolute value. Also note that $\sigma_i$ is a function of $g_i$ according to experimental observations~\cite{feinberg2018making}.

Therefore, when programming a weight, the actual value mapped onto the device, denoted as $\mathcal{W}_{\text{map}}$, would be as follows:
\begin{align}
\begin{split}
    \mathcal{W}_{map}   & = \sum_{i=0}^{M/K -1}2^{i\times K}{\mathcal{N}\left( g_i, \sigma_i^2\right) } \\
                        & = \sum_{i=0}^{M/K -1}2^{i\times K}\mathcal{N}\left(\sum_{j=0}^{K -1}{ m_{i\times K + j} \times 2^{j}, \sigma_i^2 }\right) \\
                        & = \mathcal{W}_{des} + \sum_{i=0}^{M/K -1}2^{i\times K}{\mathcal{N}(0, \sigma_i^2) }\\
                        & = \mathcal{W}_{des} +\mathcal{N}\left(0, \sum_{i=0}^{M/K-1}{\sigma_i^2 \times 2^{i\times K \times 2}}\right) \label{eq:noise}
\end{split}
\end{align}

For the simplicity of representation, we denote the relation from $g_i$ to $\sigma_i$ as 
\begin{equation}
    \sigma_i = \sigma \times \beta \times \mathcal{D}m(g_i)
\end{equation}
where $\mathcal{D}m$ is the device variation mapping function that represents different levels of device variation when the device is programmed to different values. $\beta$ is a normalization factor. \yadd{In this paper, we conduct experiments on two categories of devices: devices with uniform variations and devices with non-uniform variations.} For a device with uniform variations where $\mathcal{D}m(g_i) \equiv 1$, $\beta = 1$, \yadd{we abstract and simplify a device model from the device variation data of a representative RRAM device~\cite{yao2020fully}. The device model indicates that $\sigma = 0.1$.}
\yrpl{
For other devices, \emph{i.e.}, devices with non-uniform variations, the $\beta$ value is tuned experimentally. The goal is to ensure that with a different $\mathcal{D}m(g_i)$ function, when $\sigma = 0.1$, the DNN model has the same average performance under device variations as using devices with uniform variations.
}{
For devices with non-uniform variations, $\mathcal{D}m(g_i)$ has $2^K$ distinct values. We perform device model abstraction and simplification on two representative devices, one is a FeFET device~\cite{wei2022switching} and the other is an RRAM device~\cite{liu2023architecture}. We name them F2 and R4, respectively. We also extrapolate the modeling data to obtain a synthesized F6 device. The detailed modeling result is described in Section~\ref{sect:nu-dev}.
}

In the subsequent experiments, we adhere to the parameter setting of $K=2$ as specified in~\cite{jiang2020device} and employ the above-described model to simulate the write-verify process. Consistent with the standard procedures outlined in Section~\ref{sec:2.1}, each weight is iteratively programmed to reduce the discrepancy between the device's current value and the desired value to below 0.06. Due to inherent randomness, the number of cycles required for write-verify can differ from weight to weight. Some may not need any adjustments, while others may require several cycles. Statistically, our model yields an average of 10 cycles across all weights, resulting in a weight variation distribution with a standard deviation $\sigma=0.03$ after the write-verify process. These results align with those reported in \cite{shim2020two}, thereby confirming the validity of our model and the chosen parameters.

\subsection{Baselines and Metrics}
In addition to the common practice of exhaustive write-verify that performs write-verify on all weights offering a straightforward basis for comparison, we evaluate \USWIM~against three baselines: (1) \emph{Random selection}: each time we randomly choose a subset of weights that have not yet undergone write-verify and apply the process to them. (2) \emph {Magnitude-based approach:} We sort all the weights by their magnitudes and prioritize write-verifying the largest ones first. (3) \emph {In-situ training:} retrain the networks on-chip following the same method as that used in~\cite{yao2020fully}. No write-verify is performed. \yadd{We do not compare \USWIM~with hardware-aware training methods~\cite{rasch2023hardware, yan2023improving} and methods that accelerates the write-verify process~\cite{buchel2022gradient, mackin2022optimised} because \USWIM~is orthogonal to these methods and thus they can be used together to offer a even better solution.}
 
To ensure a fair comparison across methods using different programming mechanisms (write-verify versus on-device training), we use the total number of write cycles as the metric for evaluating programming time. This approach is justifiable as writing to NVM devices is considerably more time-consuming than reading or other operations. For the two write-verify baselines and \USWIM, the model described in Section~\ref{sect:model} is utilized to simulate and count the total number of write cycles. In the case of in-situ training, the number of writes in each iteration corresponds to the number of weights chosen for updating in that iteration, since no write-verify process is involved. To better compare different methods, we normalize the number of write cycles with respect to that used to write-verify all the weights in the DNN model under the same setting. This metric is named normalized write cycles (NWC).

Note that for \USWIM, random selection, and magnitude-based selection, $0.0 \leq$ NWC $\leq 1.0 $, because NWC $= 0.0$ means no write-verify and NWC $= 1.0$ corresponds to the conventional method of writing-verifying all the weights. For in-situ training, NWC $= 0.0$ means no in-situ training, but NWC can exceed $1.0$ because the model can be trained for many iterations and needs a large number of writes to update the weights. Specifically, as discussed in Section~\ref{sect:model}, we model the average number of write-verify cycles for each device to be 10 and this means training the whole model for 10 iterations is equal to NWC $= 1.0$.\ydel{.} In this study, we modify the maximum allowed accuracy drop, denoted as $\delta A$, for each technique, and record the resulting NWC required. 

Each model presented is quantized to the appropriate data precision. They are trained to reach convergence (\emph{i.e.}, reaching the highest possible test accuracy) on GPUs prior to being mapped to nvCiM accelerators. While the training is aware of quantization, as referenced in~\cite{jiang2020device}, it doesn't account for device variations. Our tests are performed on GTX Titan-XP GPUs using the PyTorch 1.12 machine learning framework\footnote{It's worth noting that PyTorch versions greater than or equal to 1.8, but less than 2.0 (excluding 1.10), yield results consistent with those reported in this manuscript.}. Given the inherent variability in device variations, all results presented in this document are derived from 3,000 Monte Carlo iterations, ensuring convergence. Both the average and standard deviation values are provided.

\begin{table*}[t]
    \centering
    \caption{
    Comparison of Accuracy (\%) and Normalized Write Cycles (NWC) between \USWIM~and baseline methods on LeNet for the MNIST dataset, varying by standard deviation ($\sigma$) as specified in Eq.~\ref{eq:noise} prior to write-verify. Data are aggregated from 3,000 Monte Carlo simulations and presented in a mean$\pm$standard deviation format. Write cycles are scaled relative to the cycles needed for complete write-verify. An NWC value of $0.0$ indicates no write-verify or in-situ training. An NWC value of $1.0$ for the three write-verify approaches signifies the conventional practice of write-verifying all weights.
    }
    \begin{tabular}{clccccccc}
        \toprule
        \multirow{2}{*}{$\sigma$}     & \multirow{2}{*}{Method} &\multicolumn{7}{c}{Normalized Write Cycles (NWC)}\\
       &          &  0.0 & 0.1 & 0.3 & 0.5 & 0.7 & 0.9 & 1.0 \\
        \midrule
    \multirow{4}{*}{0.1}
    & \USWIM      & $\uparrow$    & \textbf{98.49 $\pm$ 0.08} & \textbf{98.56 $\pm$ 0.08} & \textbf{98.57 $\pm$ 0.08} & \textbf{98.57 $\pm$ 0.08} & \textbf{98.57 $\pm$ 0.08} & $\uparrow$\\
    & Magnitude & \multirow{2}{*}{97.96 $\pm$ 0.31} & 98.20 $\pm$ 0.19 & 98.41 $\pm$ 0.12 & 98.50 $\pm$ 0.09 & 98.54 $\pm$ 0.08 & 98.56 $\pm$ 0.08 & 98.58 $\pm$ 0.08\\
    & Random    &               & 98.03 $\pm$ 0.26 & 98.17 $\pm$ 0.21 & 98.30 $\pm$ 0.16 & 98.42 $\pm$ 0.12 & 98.52 $\pm$ 0.09 & $\downarrow$\\
    & In-situ   & $\downarrow$  & 98.39 $\pm$ 0.21 & 98.46 $\pm$ 0.19 & 98.47 $\pm$ 0.17 & 98.48 $\pm$ 0.16 & 98.50 $\pm$ 0.17 & 98.51 $\pm$ 0.17\\
        \midrule
    \multirow{4}{*}{0.15}
    & \USWIM     & $\uparrow$    & \textbf{98.30 $\pm$ 0.13} & \textbf{98.52 $\pm$ 0.09} & \textbf{98.57 $\pm$ 0.08} & \textbf{98.57 $\pm$ 0.08} & \textbf{98.58 $\pm$ 0.08} & $\uparrow$\\
    & Magnitude & \multirow{2}{*}{96.13 $\pm$ 1.23} & 97.33 $\pm$ 0.56 & 98.14 $\pm$ 0.21 & 98.43 $\pm$ 0.12 & 98.51 $\pm$ 0.10 & 98.56 $\pm$ 0.08 & 98.58 $\pm$ 0.08\\
    & Random    &               & 96.53 $\pm$ 1.04 & 97.20 $\pm$ 0.65 & 97.73 $\pm$ 0.39 & 98.12 $\pm$ 0.23 & 98.45 $\pm$ 0.12 & $\downarrow$\\
    & In-situ   & $\downarrow$  & 96.47 $\pm$ 1.00 & 96.59 $\pm$ 0.82 & 96.69 $\pm$ 0.84 & 96.72 $\pm$ 0.82 & 96.79 $\pm$ 0.85 & 96.84 $\pm$ 0.77 \\
        \midrule
    \multirow{4}{*}{0.2}
    & \USWIM      & $\uparrow$    & \textbf{98.12 $\pm$ 0.16} & \textbf{98.46 $\pm$ 0.09} & \textbf{98.55 $\pm$ 0.08} & \textbf{98.57 $\pm$ 0.08} & \textbf{98.58 $\pm$ 0.08} & $\uparrow$\\
    & Magnitude & \multirow{2}{*}{94.46 $\pm$ 2.16} & 96.20 $\pm$ 1.11 & 97.65 $\pm$ 0.39 & 98.29 $\pm$ 0.14 & 98.45 $\pm$ 0.10 & 98.54 $\pm$ 0.08 & 98.58 $\pm$ 0.08\\
    & Random    &               & 94.89 $\pm$ 1.90 & 96.13 $\pm$ 1.20 & 97.15 $\pm$ 1.43 & 97.88 $\pm$ 0.71 & 98.38 $\pm$ 0.20 & $\downarrow$\\
    & In-situ   & $\downarrow$  & 95.33 $\pm$ 1.75 & 95.96 $\pm$ 1.36 & 96.42 $\pm$ 1.18 & 96.49 $\pm$ 1.09 & 96.69 $\pm$ 0.94 & 96.82 $\pm$ 0.80\\
    
        \bottomrule
    \end{tabular}
    \label{tab:LeNet}
\end{table*}

\begin{figure}[ht]
\centering
\subfigure[ConvNet for CIFAR-10]{
    \includegraphics[trim=10 0 30 30, clip, width=0.8\linewidth]{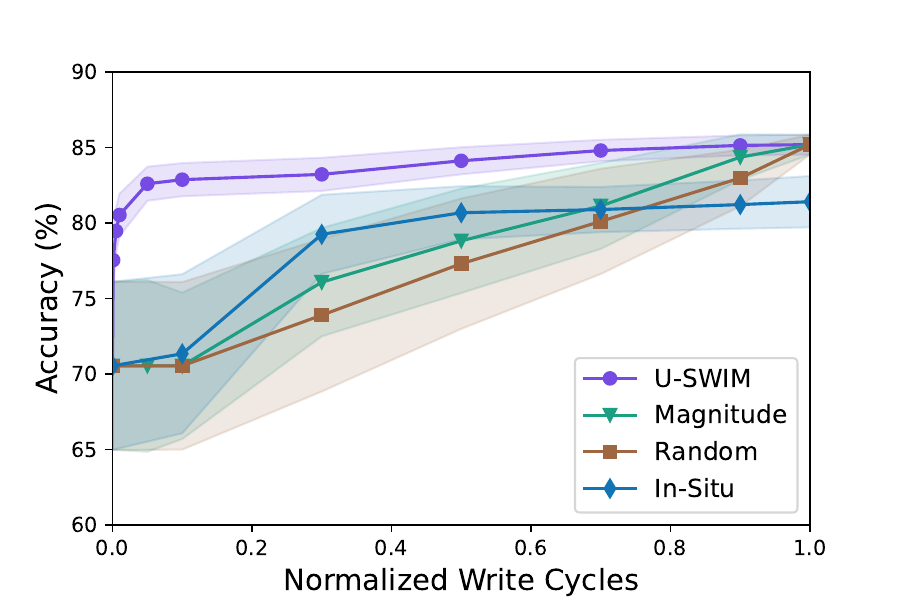}
    \label{fig:ConvNet}
}\\
\subfigure[ResNet-18 for CIFAR-10]{
    \includegraphics[trim=8 0 32 30, clip, width=0.8\linewidth]{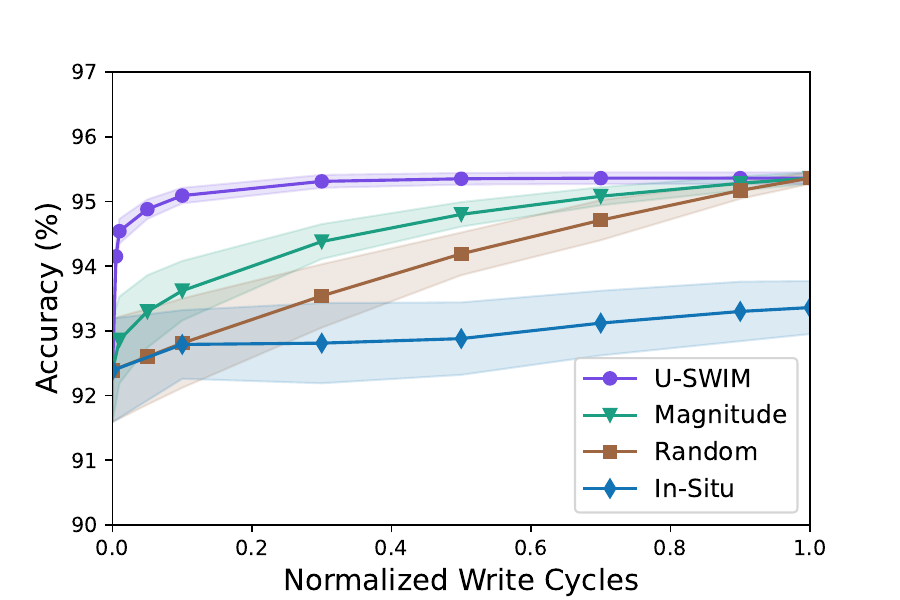}    
    \label{fig:Res18}
}\\
\subfigure[ResNet-18 for Tiny ImageNet]{
    \includegraphics[trim=10 0 30 30, clip, width=0.8\linewidth]{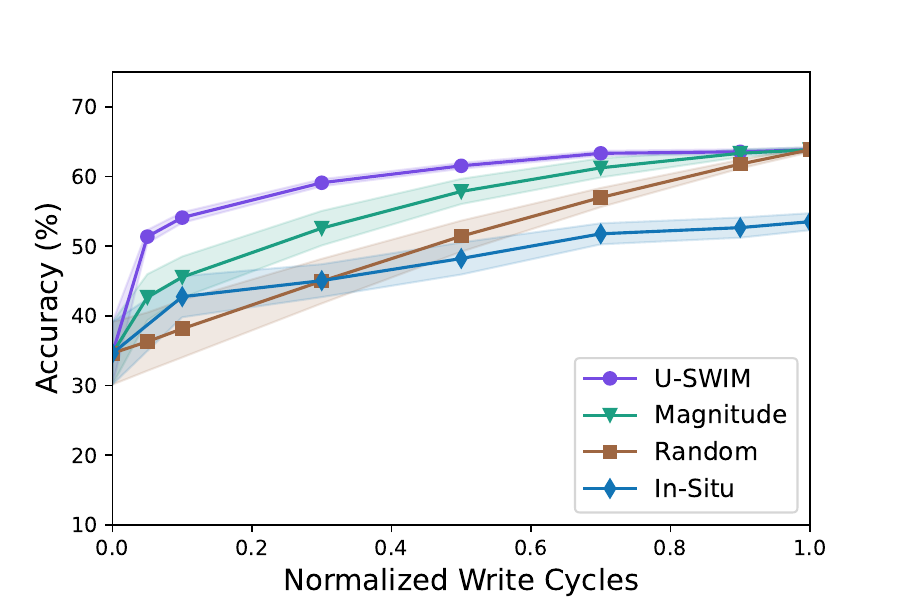}
    \label{fig:TIN}
}

\caption{
Uniform Device Variations: Comparison of Accuracy vs. Normalized Write Cycles (NWC) for \USWIM~and baseline methods across three models, ConvNet and ResNet-18 for CIFAR-10, and ResNet-18 for Tiny ImageNet. Solid lines denote mean accuracy, while shaded regions indicate standard deviation, derived from 3,000 Monte Carlo simulations utilizing the device variation model.
}
\label{fig:all_result}
\end{figure}

\subsection{Results for MNIST}

We begin our discussion by demonstrating the effectiveness of the \USWIM~approach using the LeNet model architecture applied to the MNIST dataset. For this experiment, both the weights and activations are quantized to 4-bit representations. In the scenario free of device variations, the model achieves an accuracy rate of 98.68\%. The model contains a total of $1.05 \times 10^5$ weights.

While commonly devices usually display a standard deviation ($\sigma$) of 0.1 with respect to device variation prior to the write-verify process. However, emerging technologies still in their developmental stages can exhibit more substantial variations. To establish the broad applicability of \USWIM, we evaluate its performance relative to other baseline approaches at different $\sigma$ values, as summarized in Table~\ref{tab:LeNet}. The data reveals that complete write-verification of all weights effectively recovers the model's initial accuracy, reaching a score of 98.58\% when the normalized write cycle (NWC) is set at 1.0 across all write-verify techniques. Even though a decline in accuracy is observed as NWC diminishes for all methods, \USWIM~consistently needs fewer NWCs to sustain similar accuracy levels across diverse $\sigma$ conditions. Additionally, \USWIM~demonstrates a significantly smaller standard deviation in accuracy over the course of 3,000 Monte Carlo trials, indicating that it would likely deliver highly consistent performance across various devices.

In specific terms, when compared to the traditional method of write-verify for every weight (NWC $=1.0$), \USWIM~requires only half of the write cycles (NWC $=0.5$, or a $2\times$ speedup) under typical variations ($\sigma=0.1$) to maintain the original accuracy. Even at a mere 10\% of the write cycles (NWC $=0.1$ or $10\times$ speedup), the accuracy decline using \USWIM~is less than $0.1\%$. In contrast, the magnitude-based strategy, the random method, and in-situ training require an NWC of approximately 0.5, 0.9, and 0.9, respectively, to achieve the same accuracy level. This results in a speedup of $5\times$, $9\times$, and $9\times$ for \USWIM~relative to these methods. Additionally, \USWIM~continues to perform well even when $\sigma$ increases to $0.2$. Using only 10\% of the write cycles, it experiences an accuracy reduction of less than 0.5\%. For the same level of accuracy, the random and magnitude-based approaches demand 70\% and 50\% of the write cycles, respectively. Meanwhile, in-situ training, even with $10\times$ the write cycles (NWC $=1.0$), fails to match this accuracy, suggesting that additional training iterations are still needed. Although not displayed in Table~\ref{tab:LeNet}, in-situ training can fully restore the model's accuracy to 98.68\% using 32 NWCs. While this suggests it could outperform the write-verify methods in terms of accuracy, it does so at the expense of significantly more writes, extended programming time, and the requirement of extra hardware.

\subsection{Results for CIFAR-10}

We proceed to evaluate the performance of \USWIM~using the CIFAR-10 dataset on two DNN architectures: ConvNet~\cite{peng2019dnn+} and ResNet-18~\cite{he2016deep}. For both models, the weights and activations are quantized to 6 bits, and the standard deviation $\sigma$ is set to 0.1 prior to the write-verify process. In the absence of device variations, ConvNet and ResNet-18 achieve accuracies of 86.07\% and 95.62\%, respectively. When device variations are accounted for and all weights are subjected to write-verify, the accuracies change to 85.19\% for ConvNet and 95.36\% for ResNet-18. The total counts of weights for ConvNet and ResNet-18 are $6.40 \times 10^6$ and $1.12 \times 10^7$, respectively.

As shown in Fig.~\ref{fig:ConvNet}, a comparison of \USWIM~with other baseline methods on ConvNet reveals that, except for \USWIM, all methods experience an accuracy reduction of over 10\% when the NWC is set to 0.1. In contrast, \USWIM~limits this drop to below 2.5\%. Notably, \USWIM~also exhibits the smallest standard deviation in accuracy, highlighting its robustness. Although not displayed in Fig.~\ref{fig:ConvNet}, in-situ training can fully recover the model's accuracy with an NWC of 75.

Similarly, Fig.~\ref{fig:Res18} demonstrates the comparative efficiency of \USWIM~against baseline methods on ResNet-18. As with ConvNet, \USWIM~outperforms the other methods by keeping the accuracy drop below 0.5\% with only 10\% of the write cycles. The other methods, in the same setting, suffer from an accuracy reduction greater than 2\%. In-situ training is capable of restoring full model accuracy but requires 115 NWCs to do so.

\subsection{Experiments on Tiny ImageNet}

Lastly, we evaluate the efficacy of \USWIM~using the Tiny ImageNet dataset with the ResNet-18 architecture~\cite{he2016deep}, adhering to the same quantization and $\sigma$ settings. Without device variations, the model achieves an accuracy of 65.50\%. With device variations considered and all weights verified through writing, the accuracy slightly dips to 64.84\%. The model has a total of $1.13 \times 10^7$ weights.

Fig.~\ref{fig:TIN} compares the performance of \USWIM~with other baseline methods on ResNet-18 for Tiny ImageNet. Given that Tiny ImageNet presents a more formidable challenge compared to CIFAR-10, all methods exhibit more substantial accuracy declines, as evidenced by comparison with Fig.~\ref{fig:Res18}. Despite this, \USWIM~manages to limit the accuracy decrease to less than 3\% when utilizing just 10\% of the write cycles—the best performance among all tested methods. In-situ training can achieve full recovery of model accuracy, but it requires as many as 155 NWCs to do so.

\begin{figure}
\centering
\subfigure[ConvNet for CIFAR-10]{
    \includegraphics[trim=8 0 32 30, clip, width=0.8\linewidth]{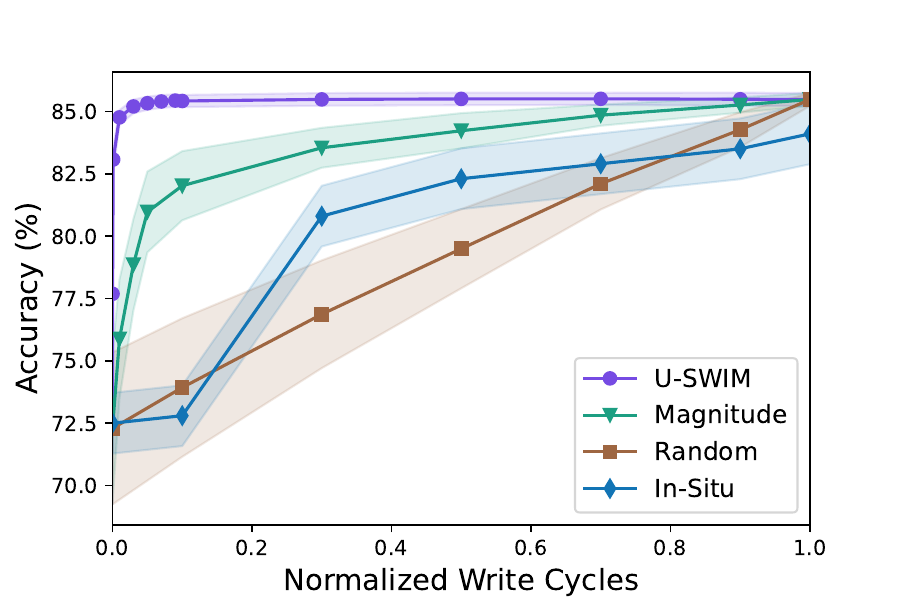}
    \label{fig:ConvNet_4}
}\\
\subfigure[ResNet-18 for CIFAR-10]{
    \includegraphics[trim=8 0 32 30, clip, width=0.8\linewidth]{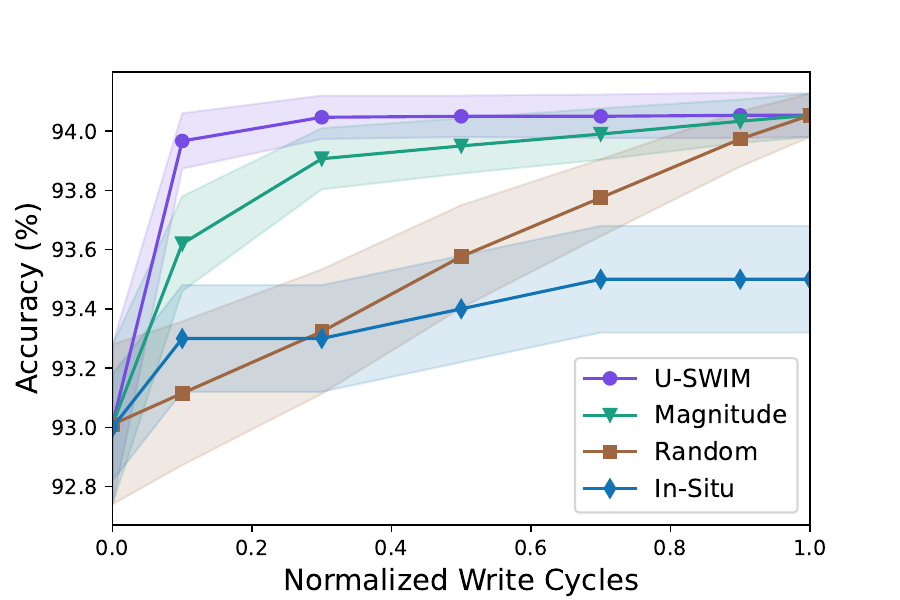}
    \label{fig:Res18_4}
}\\
\subfigure[ResNet-18 for Tiny ImageNet]{
    \includegraphics[trim=10 0 30 30, clip, width=0.8\linewidth]{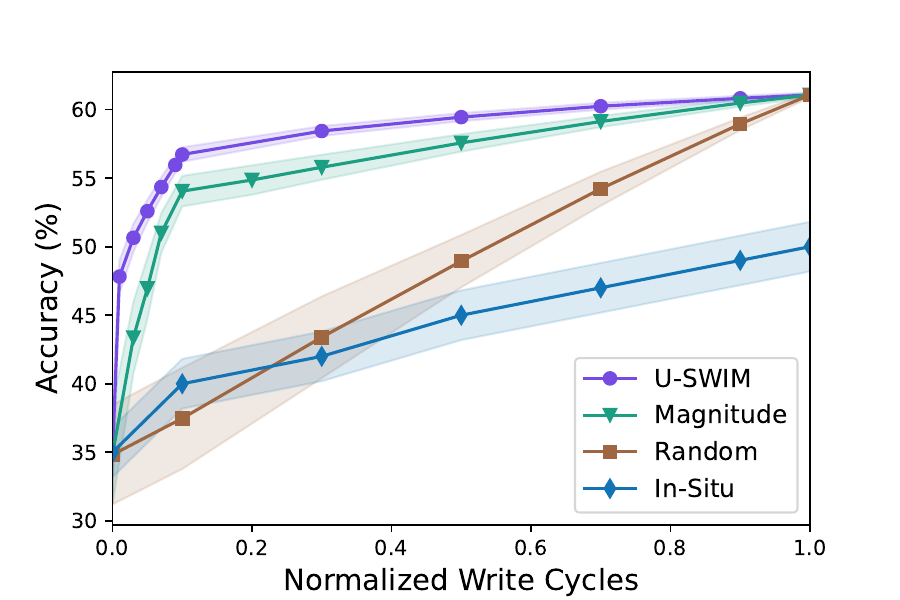}
    \label{fig:TIN_4}
}

\caption{
Non-Uniform Device Variations: Comparison of Accuracy vs. Normalized Write Cycles (NWC) for \USWIM~and baseline methods across three models, ConvNet and ResNet-18 for CIFAR-10, and ResNet-18 for Tiny ImageNet—using Device \yrpl{$F_4$}{$R_4$}. Solid lines denote mean accuracy, while shaded regions indicate standard deviation. These metrics are based on 3,000 Monte Carlo simulations using the specified device variation model.
}
\label{fig:all_result_4}
\end{figure}

\subsection{Description for Devices with Non-uniform Variations}\label{sect:nu-dev}
We've shown the effectiveness of \USWIM~for RRAM devices with uniform variations (\emph{i.e.}, $\sigma_i \equiv 1$ regardless of $g_i$ values) in previous experiments. In this section, we show the effectiveness of \USWIM~on devices with non-uniform variations. We consider three different types, each having four levels. This means (1) each device is able to represent $g_i$ values of $0$, $1$, $2$, and $3$ and (2) $\sigma_i$ differs when $g_i$ changes. 

Specifically, we consider the first device to be a FeFET device named $F_2$ having a $\sigma_i$-$g_i$ relationship of
\begin{equation}
    \sigma_{iF_2} = \begin{cases}
            1,\ \ \  if\  g_i = 0\\
            2,\ \ \  if\  g_i = 1\\
            2,\ \ \  if\  g_i = 2\\
            1,\ \ \  if\  g_i = 3\\
            \end{cases}\label{eq:f2}
\end{equation}
and have a corresponding $\beta$ value of 0.8. \yadd{This device model is abstracted from data reported in~\cite{wei2022switching}.}

We then consider the second device to be an RRAM device named $R_4$ having a $\sigma_i$-$g_i$ relationship of
\begin{equation}
    \sigma_{iR_4} = \begin{cases}
            1,\ \ \  if\  g_i = 0\\
            4,\ \ \  if\  g_i = 1\\
            4,\ \ \  if\  g_i = 2\\
            1,\ \ \  if\  g_i = 3\\
            \end{cases}\label{eq:r4}
\end{equation}
and have a corresponding $\beta$ value of 0.57. \yadd{This device model is abstracted from data reported in~\cite{liu2023architecture}.}

We finally consider the third device to be a FeFET device named $F_6$ having a $\sigma_i$-$g_i$ relationship of
\begin{equation}
    \sigma_{iF_6} = \begin{cases}
            1,\ \ \  if\  g_i = 0\\
            6,\ \ \  if\  g_i = 1\\
            6,\ \ \  if\  g_i = 2\\
            1,\ \ \  if\  g_i = 3\\
            \end{cases}\label{eq:f6}
\end{equation}
and have a corresponding $\beta$ value of 0.43. \yadd{This device model is extrapolated from the device data of the two aforementioned devices.}

We present the experimental results of performing selective write-verify on nvCiM DNN accelerators with these three devices. To demonstrate \USWIM's effectiveness, we compare the average performance of models utilizing \USWIM~as the selective write-verify metric against the baselines the same as earlier experiments.

\begin{figure}
\centering
\subfigure[ConvNet for CIFAR-10 on device $F_2$]{
    \includegraphics[trim=8 0 32 30, clip, width=0.8\linewidth]{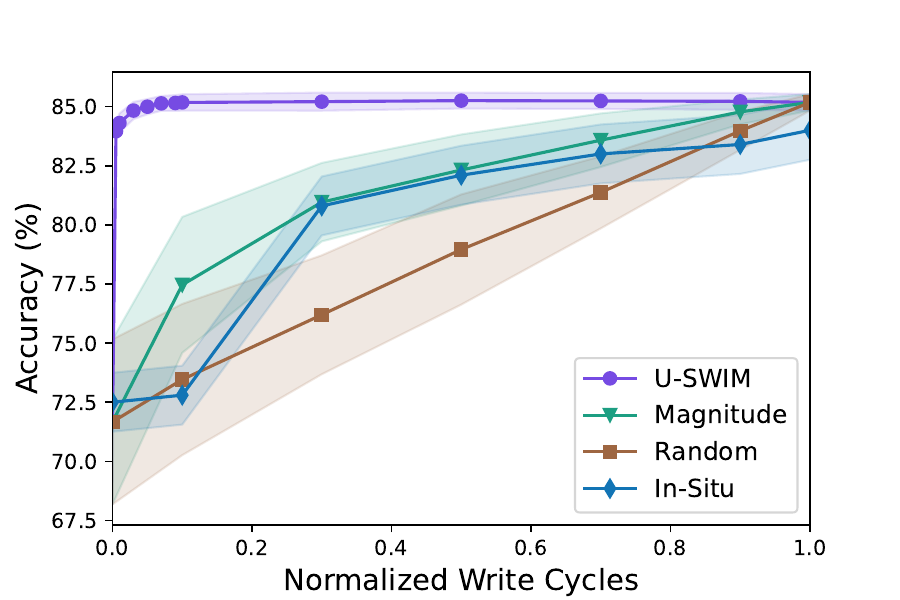}
    \label{fig:S2}
}\\
\subfigure[ConvNet for CIFAR-10 on device \yrpl{$F_4$}{$R_4$}]{
    \includegraphics[trim=8 0 30 30, clip, width=0.8\linewidth]{figures/QCIFAR_S4D0.15.pdf}
    \label{fig:S4}
}\\
\subfigure[ConvNet for CIFAR-10 on device $F_6$]{
    \includegraphics[trim=8 0 32 30, clip, width=0.8\linewidth]{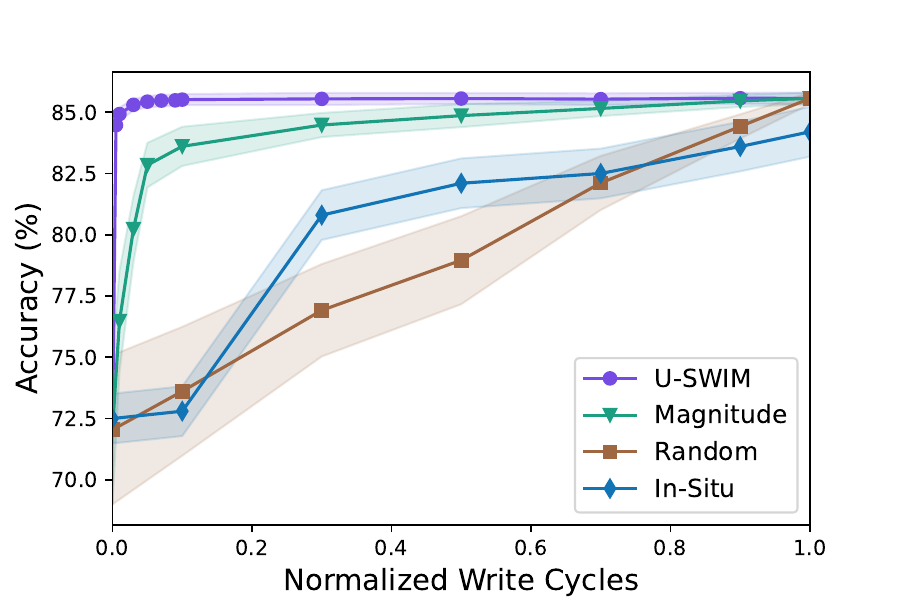}
    \label{fig:S6}
}

\caption{Accuracy vs. Normalized Write Cycles (NWC) in ConvNet for CIFAR-10: Comparison between \USWIM~and baseline methods using three different devices, $F_2$, $R_4$, and $F_6$. Solid lines indicate average accuracy, while shaded regions represent the standard deviation. These results are obtained from 3,000 Monte Carlo simulations using the respective device variation models.
}
\label{fig:diff_s}
\end{figure}

\begin{table*}[t]
    \centering
    \caption{Ablation Study: A comparison between the original SWIM and its extended version, \USWIM, on ConvNet for CIFAR-10 is presented. The evaluation focuses on various types of devices, specified by the standard deviation in Eq.~\ref{eq:noise} prior to the write-verify process. Data are collected from 3,000 Monte Carlo simulations and reported in a mean$\pm$std format. Write cycles are normalized relative to the cycles required to write-verify all weights. An NWC of $0.0$ implies no write-verify or in-situ training was used, while an NWC of $1.0$ for the three write-verify methods aligns with the conventional practice of write-verifying all weights.
    }
    \begin{tabular}{clccccccc}
        \toprule
        \multirow{2}{*}{Device}     & \multirow{2}{*}{Method} &\multicolumn{7}{c}{Normalized Write Cycles (NWC)}\\
       &          &  0.0 & 0.1 & 0.3 & 0.5 & 0.7 & 0.9 & 1.0 \\
        \midrule
    \multirow{3}{*}{F2}
    & \USWIM & \multirow{3}{*}{71.67 $\pm$ 3.49} & 85.18 $\pm$ 0.35 & 85.21 $\pm$ 0.38 & 85.25 $\pm$ 0.34 & 85.24 $\pm$ 0.34 & 85.22 $\pm$ 0.36 & \multirow{3}{*}{85.18 $\pm$ 0.35} \\
    & \emph{SWIM} &                                   & 82.89 $\pm$ 0.65 & 84.11 $\pm$ 0.50 & 84.67 $\pm$ 0.41 & 85.02 $\pm$ 0.36 & 85.22 $\pm$ 0.34 & \\
    & Magnitude    &                                   & 77.46 $\pm$ 2.87 & 80.96 $\pm$ 1.66 & 82.32 $\pm$ 1.50 & 83.58 $\pm$ 1.13 & 84.78 $\pm$ 0.52 & \\
        \midrule
    \multirow{3}{*}{R4}
    & \USWIM   & \multirow{3}{*}{72.29 $\pm$ 3.05} & 85.42 $\pm$ 0.24 & 85.48 $\pm$ 0.25 & 85.50 $\pm$ 0.25 & 85.50 $\pm$ 0.24 & 85.49 $\pm$ 0.26 & \multirow{3}{*}{85.48 $\pm$ 0.26} \\
    & \emph{SWIM}  &                                   & 84.36 $\pm$ 0.38 & 84.94 $\pm$ 0.35 & 85.22 $\pm$ 0.28 & 85.39 $\pm$ 0.27 & 85.46 $\pm$ 0.27 & \\
    & Magnitude     &               & 82.02 $\pm$ 1.39 & 83.54 $\pm$ 0.80 & 84.22 $\pm$ 0.71 & 84.85 $\pm$ 0.42 & 85.26 $\pm$ 0.31 & \\
    \midrule
    \multirow{3}{*}{F6}
    & \USWIM   & \multirow{3}{*}{72.04 $\pm$ 3.05} & 85.51 $\pm$ 0.23 & 85.54 $\pm$ 0.25 & 85.56 $\pm$ 0.23 & 85.53 $\pm$ 0.24 & 85.57 $\pm$ 0.22 & \multirow{3}{*}{85.56 $\pm$ 0.24} \\ 
    & \emph{SWIM}  &                                   & 84.93 $\pm$ 0.32 & 85.23 $\pm$ 0.26 & 85.40 $\pm$ 0.23 & 85.48 $\pm$ 0.23 & 85.54 $\pm$ 0.25 & \\
    & Magnitude    &                                    & 83.61 $\pm$ 0.80 & 84.48 $\pm$ 0.49 & 84.86 $\pm$ 0.47 & 85.15 $\pm$ 0.31 & 85.46 $\pm$ 0.25 & \\
    \bottomrule
    \end{tabular}
    \label{tab:SvU}
\end{table*}

\subsection{Effective of USWIM for Different Models under Non-uniform Variations}
We first demonstrate the effectiveness of \USWIM~on the previously used CIFAR-10 and Tiny ImageNet dataset with the two previously used models ConvNet~\cite{peng2019dnn+} and ResNet-18~\cite{he2016deep} with the same settings and the same set of weight values as previously used. 

Fig.~\ref{fig:ConvNet_4} presents a comparison of \USWIM~against the baseline methods on ConvNets targeting the CIFAR-10 dataset. When all weights undergo the write-verify process, every method, with the exception of \USWIM, registers an accuracy decrease exceeding 7\% at an NWC of 0.1. In contrast, \USWIM~manages to restrict this reduction to under 1.2\%. This visualization clearly illustrates that \USWIM~offers the least variance in accuracy, showcasing its remarkable resilience. It's worth noting, although not depicted in Fig.~\ref{fig:ConvNet_4}, that at an NWC of 86, in-situ training is capable of restoring the model's accuracy fully.

Moving forward, Fig.~\ref{fig:Res18_4} compares \USWIM~with the baseline methods for ResNet-18 targeting the CIFAR-10 dataset. The inferences here mirror previous observations. When compared to the process of writing and verifying all weights, \USWIM~is capable of limiting the accuracy drop to below 0.2\%, even when utilizing just 10\% of the write cycles. On the other hand, alternative methods see an accuracy drop of over 0.5\% for the equivalent number of write cycles. Notably, in-situ training achieves full accuracy restoration at 118 NWC.

Lastly, Fig.~\ref{fig:TIN_4} pits \USWIM~against the baseline methods using ResNet-18 on the Tiny ImageNet dataset. Given the greater complexity of this task compared to CIFAR-10, it's evident that the accuracy reductions for all methods are more significant than those observed in Fig.~\ref{fig:Res18_4}. Nevertheless, \USWIM~attains an accuracy that is less than 4\% below the standard write-verify process for all weights, even when leveraging merely 10\% of the write cycles—a feat unmatched by other methods. Impressively, in-situ training manages to recover the full model accuracy at 159 NWC.

\subsection{Effective of USWIM for Different Devices under Non-uniform Variations}\label{sect:diff_s}
We then show the effectiveness of \USWIM~on the three aforementioned devices using the previously used CIFAR-10 dataset with the ConvNet model with the same settings and the same set of weight values as previously used.

As shown in Fig.~\ref{fig:diff_s}, \USWIM~is consistently effective in reducing the portion of devices that require write-verify. When NWC = 0.01, the deployed model reaches an accuracy degradation of 0.5\% compared with writing-verifying all devices. This means 90x, 70x, and 50x improvement over using weight magnitude as a metric and over 100x improvement over random selection.

Different from \USWIM~and random selection which show consistent trends across different devices, using weight magnitude as the selection metric yields considerably different results when using different devices. Specifically, it produces increasingly better results as the difference in $\sigma_i$ over different $g_i$ values increases. Observing Eq.~\ref{eq:f2}-\ref{eq:f6} and recalling the \USWIM~metric $\frac{\partial^2 f}{\partial \tilde{w}_i^2} E[(\Delta w_i)^2]$, it becomes evident that when the difference between $\sigma_i$ increases, $E[(\Delta w_i)^2]$ plays a more important role in this metric. Empirical results indicate that the weight distribution is Gaussian. The weight magnitude metric first selects a small portion of the weights in the tail of the Gaussian distribution with high magnitudes that $g_i = 3$, and these weights have small $E[(\Delta w_i)^2]$. This is why the weight magnitude metric is ineffective when NWC is very small. Subsequently, it begins to select weights with $g_i = 2$ and then $g_i = 1$, which has a large $E[(\Delta w_i)^2]$, thus becoming more effective. However, the weight magnitude method does not take the $\frac{\partial^2 f}{\partial \tilde{w}_i^2}$ factor into account so it is consistently inferior to \USWIM.

\subsection{Comparison with the Original SWIM Method}
Here we perform an ablation study to show the superiority of the proposed \USWIM~method over the original {\em SWIM} method that does not take non-uniform device variations into account. In terms of implementation, the original {\em SWIM} utilizes only the second derivatives of each weight value as a sensitivity metric and fails to consider the influence of differing weight values.
This would be suitable for devices with uniform variations, where the device value deviation is not affected by its target device value. This situation serves as a special case of \USWIM~meaning that \USWIM~and {\em SWIM} behave identically when dealing with devices with uniform variations. However, since {\em SWIM} does not factor in the influence of target device value, it performs poorly in finding sensitive weights for nvCiM accelerators using such devices. As shown in Table~\ref{tab:SvU}, although {\em SWIM} achieves up to 5.4\% improvement in accuracy compared to the baseline that uses weight magnitude as the sensitivity metric, it still shows a lower 2.3\% lower average accuracy when compared with the extended \USWIM~method, which accounts for the non-uniform device variations into account.

\yadd{Note that the improvement of \USWIM~compared with the original {\em SWIM} method differs when the device types changes. Specifically, as the difference in device variation of different conductance values increases, the absolute improvement of the \USWIM~method over {\em SWIM} decreases. This may contradict the claim that \USWIM~is supports devices with non-uniform variations better. However, the experimental results are actually consistent with the claim because (1) the effectiveness of the original {\em SWIM} method decreases as the difference in device variations increases and (2) the relative improvement compared to the magnitude-based baseline is consistently around 50\%.}

\begin{figure}[ht]
\begin{center}
\centerline{\includegraphics[trim=30 290 350 20, clip, width=1.0\linewidth] {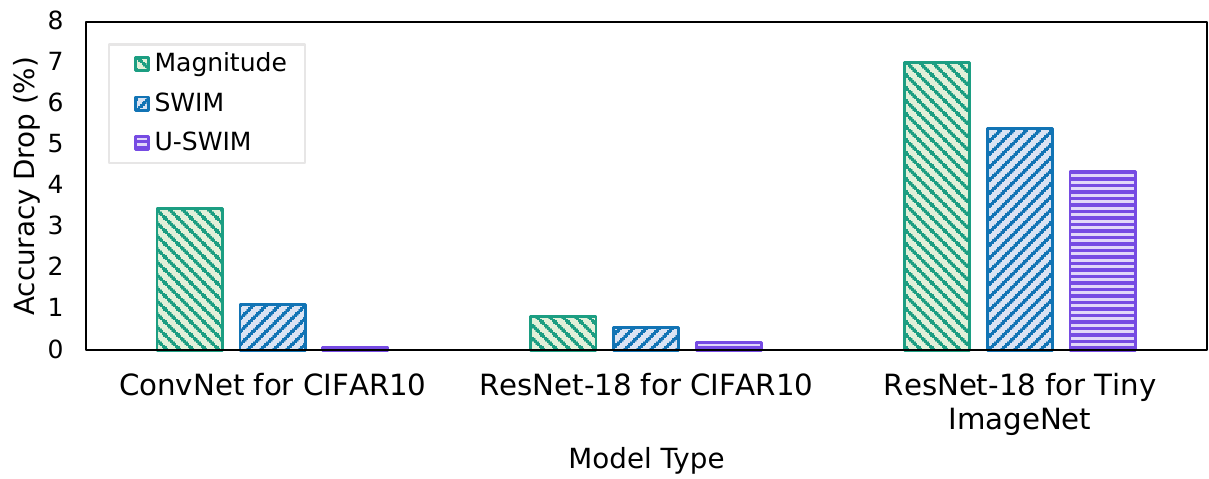}}
\caption{
\yadd{Non-Uniform Device Variations: Comparison of accuracy drops (from accuracy w/o device variations) under NWC=0.1 for \USWIM~ and {\em SWIM} method (also including the magnitude-based method as a baseline) across three models, ConvNet and ResNet-18 for CIFAR-10, and ResNet-18 for Tiny ImageNet models using Device $R_4$. These metrics are based on 3,000 Monte Carlo simulations using the specified device variation model.}}
\label{fig:SvU}
\end{center}
\end{figure}

\yadd{Similar experiments are conducted for different models to demonstrate the scalability of \USWIM~compared with the original {\em SWIM} method. Fig.~\ref{fig:SvU} presents the comparison results of \USWIM~and {\em SWIM} methods in ConvNet and ResNet-18 for CIFAR-10, and ResNet-18 for Tiny ImageNet models using device $R_4$ and with a NWC of $0.1$. In ConvNet for CIFAR-10, the {\em SWIM} method show an accuracy improvement from the magnitude-based baseline by 2\% and \USWIM~shows a further 1\% improvement, translating to a 50\% relative increase. Similar results are observed from experimental results of ResNet-18 models for CIFAR-10 and Tiny ImageNet shows a relative increase of 100\% and 40\%. Although the relative increase is significant, the absolute improvement is marginal because the influence of device variations is not that profound, \emph{i.e.}, the accuracy without write-verify is already high.}

\section{Conclusions \yadd{and Future Works}}
In this study, we challenge the prevalent approach of write-verifying every weight of a DNN when mapping it onto an nvCiM platform to address device imperfections. We demonstrate that only a fraction of the weights need to undergo the write-verify process while still retaining accuracy. This can significantly reduce the programming duration. Our newly introduced \USWIM~method efficiently calculates second derivatives to select weights for write-verify. Experimental findings indicate a speedup of up to 10x compared to traditional write-verify approaches with negligible accuracy trade-offs. Additionally, \USWIM~outperforms our earlier {\em SWIM} method, achieving a 7x speedup when handling devices with non-uniform variations.

\yadd{Despite the impressive effectiveness of \USWIM, it is yet to be tested for (1) devices with non-Gaussian distributed variations and (2) models including recurrent neural networks. The device model also overlooked the influence of write-verifying one device to other devices. These can be efficiently done by extending the \USWIM~framework and will be included in future works.}

\section{Acknowledgement}
This project is supported in part by the National Science Foundation under grants CNS-1822099 and CCF-1919167 and also by ACCESS – AI Chip Center for Emerging Smart Systems, sponsored by InnoHK funding, Hong Kong SAR and Semiconductor Research Corporation (SRC). Code is available at https://github.com/MariusAnje/SWIM

\bibliographystyle{IEEEtran}
\bibliography{M6_References}

\begin{IEEEbiography}[{\includegraphics[width=1in,height=1.25in,clip,keepaspectratio]{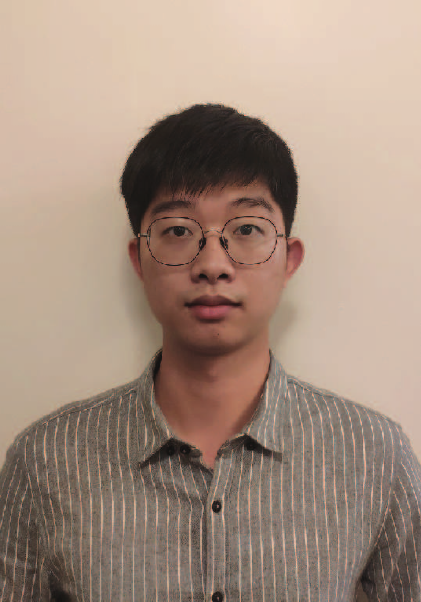}}]{Zheyu Yan} 
is a Ph.D. student in the Department of Computer Science and Engineering at the University of Notre Dame. He earned his B.S. degree from Zhejiang University in 2019. He is deeply interested in the co-design of software and hardware for deep neural network accelerators, particularly focusing on non-volatile memory-based compute-in-memory platforms.
\end{IEEEbiography}

\begin{IEEEbiography}[{\includegraphics[width=1in,height=1.25in,clip,keepaspectratio]{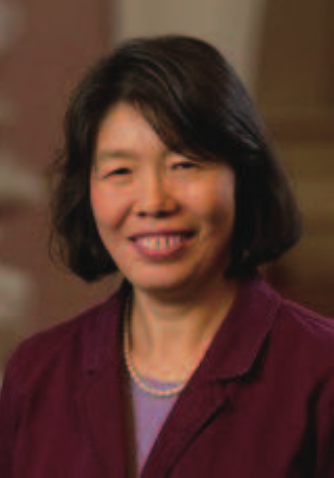}}]{Xiaobo Sharon Hu} 
earned her B.S. degree from Tianjin University in China in 1982, followed by an M.S. from the Polytechnic Institute of New York in 1984, and then a Ph.D. from Purdue University in West Lafayette, IN, USA, in 1989.

She currently holds a Professor position at the Department of Computer Science and Engineering at the University of Notre Dame, Notre Dame, IN, USA. Her research primarily revolves around beyond-CMOS technologies computing, low-power system design, and cyber-physical systems. Dr. Hu was honored with the NSF CAREER Award in 1997 and received Best Paper Awards from the Design Automation Conference in 2001 and the ACM/IEEE International Symposium on Low Power Electronics and Design in 2018.

In 2018, Dr. Hu took on the role of General Chair for the Design Automation Conference (DAC). She has been an Associate Editor for various publications including IEEE Transactions on Very Large Scale Integration (VLSI) Systems, ACM Transactions on Design Automation of Electronic Systems, and ACM Transactions on Embedded Computing. Currently, she serves as an Associate Editor for the ACM Transactions on Cyber-Physical Systems.
\end{IEEEbiography}

\begin{IEEEbiography}[{\includegraphics[width=1in,height=1.25in,clip,keepaspectratio]{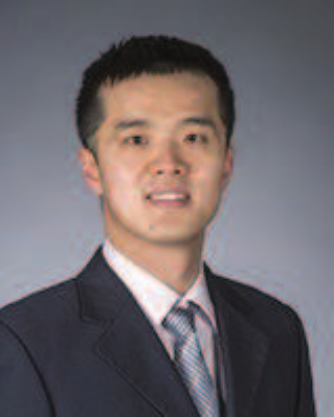}}]{Yiyu Shi} 
earned his B.S. degree with honors in electronic engineering from Tsinghua University in Beijing, China, in 2005. He later pursued his M.S. and Ph.D. degrees in electrical engineering at the University of California, Los Angeles, completing them in 2007 and 2009, respectively. He now serves as a Professor in the Departments of Computer Science and Engineering as well as Electrical Engineering at the University of Notre Dame, Notre Dame, IN, USA. His research primarily focuses on 3-D integrated circuits, hardware security, and applications in renewable energy. 

Prof. Shi has been recognized with multiple best paper nominations at premier conferences. He received the IBM Invention Achievement Award in 2009 and was honored with the Japan Society for the Promotion of Science Faculty Invitation Fellowship, the Humboldt Research Fellowship for Experienced Researchers, and the National Science Foundation CAREER Award. Additionally, he was the recipient of the IEEE Region 5 Outstanding Individual Achievement Award and the Air Force Summer Faculty Fellowship.
\end{IEEEbiography}

\end{document}